\pdfoutput=1

\documentclass[11pt,twoside,a4paper,cmspaper,final,collab]{cms-tdr}
\def\svnVersion{188638}\def\cmsCernNo{2012-282}\def\cmsCernDate{\today}\def\cmsMessage{Submitted to Physical Review D}

\begin{document}\cmsNoteHeader{EXO-11-074}

\hyphenation{had-ron-i-za-tion}
\hyphenation{cal-or-i-me-ter}
\hyphenation{de-vices}

\RCS$Revision: 182539 $
\RCS$HeadURL: svn+ssh://svn.cern.ch/reps/tdr2/papers/EXO-11-074/trunk/EXO-11-074.tex $
\RCS$Id: EXO-11-074.tex 182539 2013-04-24 09:24:30Z wulsin $
\cmsNoteHeader{EXO-11-074} 

\newcommand{\lsim}{\mathrel{\hbox{\rlap{\lower.55ex
\hbox{$\sim$}} \kern-.3em \raise.4ex \hbox{$<$}}}}
\ifthenelse{\boolean{cms@external}}{%
\newcommand{\scotchrule[1]}{\centering\begin{ruledtabular}\begin{tabular}{#1}}
\newcommand{\donescotchrule}{\end{tabular}\end{ruledtabular}}
}{
\newcommand{\scotchrule[1]}{\centering\begin{tabular}{#1}\hline\hline}
\newcommand{\donescotchrule}{\hline\hline\end{tabular}}
}
\providecommand\dedx{\ddinline{E}{x}\xspace}
\providecommand\rad{\ensuremath{\mathrm{\,rad}}\xspace}      
\providecommand\MeVcm{\ensuremath{\,\text{Me\hspace{-.08em}V\hspace{-.11em}/\hspace{-.06em}cm}}\xspace}
\newlength\cmsFigWidth
\ifthenelse{\boolean{cms@external}}{\setlength\cmsFigWidth{0.48\textwidth}}{\setlength\cmsFigWidth{0.7\linewidth}}
\title{Search for fractionally charged particles in \texorpdfstring{$\Pp\Pp$ collisions at $\sqrt{s}=7\TeV$}{pp collisions at sqrt(s) = 7 TeV}}

\date{\today}

\abstract{
A search is presented for free heavy long-lived fractionally charged
particles produced in $\Pp\Pp$ collisions at  $\sqrt{s}=7\TeV$.
The data sample was
recorded by the CMS detector at the LHC and corresponds to an
integrated luminosity of 5.0\fbinv.
Candidate fractionally charged
particles are identified by selecting tracks with associated
low charge measurements in the silicon tracking
detector. Observations are found to be consistent with expectations
for background processes.
The results of the search
are used to set upper limits on the cross section for pair
production of fractionally charged, massive spin-1/2 particles that
are neutral under $SU(3)_C$ and $SU(2)_L$.
We exclude at 95\% confidence level such particles with electric
charge  ${\pm}2e/3$ with masses below 310\GeV, and those with charge
${\pm}e/3$ with masses below 140\GeV.
}

\hypersetup{%
pdfauthor={CMS Collaboration},
pdftitle={Search for fractionally charged particles in pp collisions at sqrt(s) = 7 TeV},%
pdfsubject={CMS},%
pdfkeywords={CMS, physics, software, computing}}

\maketitle

\section{Introduction}

The reasons for expecting physics beyond the standard model to
manifest itself in $\Pp\Pp$ collisions at the Large Hadron Collider (LHC)
are as compelling as ever.
As suggested in Ref.
\cite{Meade2011}, one example of physics beyond the standard
model that may have eluded previous searches
is a new particle with an electric charge less than that of the
electron.  Owing to their lower ionization energy loss, the trajectories
of such fractionally charged particles may not pass typical track
quality requirements and a dedicated analysis is required.

While fractionally charged particles are common in some theoretical
scenarios (\eg, superstrings \cite{Wen:1985qj, Athanasiu:1988uj}), a
variety of searches for these objects in bulk matter, cosmic
rays and accelerator based experiments have reported null
results~\cite{Perl:2009zz}.
Strong constraints on models with fractionally charged
states come from astrophysics and cosmology~\cite{Langacker:2011db}.
These constraints, however, do not apply if the reheating temperature of the
universe after the last stage of inflation is much lower than the mass
of the fractionally charged particle such that there is no thermal
relic from freeze-out~\cite{Nate}.  We search for such
particles, using as a benchmark the scenario considered
in~\cite{Langacker:2011db}, namely a model with new, fractionally
charged, massive spin-1/2 particles that are neutral under $SU(3)_C$
and $SU(2)_L$ and therefore couple only to the photon and the \cPZ.  We
denote such particles with fractional charge ${\pm}qe$, where $e$ is the
charge of the electron, as $L_q$, and
assume they have a lifetime sufficiently long such that they do not
decay within the detector volume.

An interesting possibility is that these $L_q$ could also be charged
under a new asymptotically free gauge group $SU(N)$ with a confinement
scale $\Lambda$.  This would make them
a variant type of ``quirk", which are quark-like, naturally
fractionally charged particles~\cite{Kang:2008ea}.  
For $\Lambda \lsim 100$ eV, the confining string between the
quirk-antiquirk pair would have a negligible impact on their
trajectories over distances typical of collider-detector dimensions.
Thus such quirks would have the same kinematic properties and
experimental signature as the benchmark model considered in this
paper. However, the existence of 
this string would cause eventual annihilation of any pairs formed in
the early universe resulting in a negligible relic
abundance~\cite{Nate}, thereby evading the constraints
cited by Ref.~\cite{Langacker:2011db} irrespective of cosmological
history.

We search for $L_q,\overline{L}_q$ particles in a sample of tracks with a
muon-like signature.  We identify fractionally charged particle
candidates by their anomalously low ionization energy loss in the
inner tracker.  This study complements Compact Muon Solenoid (CMS)
searches for heavy stable
charged particles with anomalously high ionization energy loss
\cite{HSCP2010, HSCP2011}.

\section{Signal simulation}
Pair production of $L_q \overline{L}_q$ at the LHC proceeds via a modified
Drell--Yan mechanism with weak isospin $t_{3L}=0$, which has $L_q$-\cPZ\
axial coupling $g_A= 0$ and
vector coupling $g_V = -2 q \sin^2 \theta_W$.
We have performed Monte Carlo simulations of this signal using \PYTHIA
v6.422 \cite{Sjostrand:2006za}, with $q= 1/3, 2/3$, and $1$, and with
masses of 100, 200, 300, 400, 500, and 600\GeV.  The cross sections
are calculated to leading order with the CTEQ6L1 parton distribution
functions.
The detector response is modeled with a simulation based on
\GEANTfour~\cite{GEANT4}.

\section{Detector and data sample}
The central feature of the CMS
apparatus~\cite{CMS:2008zzk} is a
superconducting solenoid of 6\unit{m} internal diameter, providing a
magnetic field of 3.8\unit{T}. Within the field volume are a silicon
pixel and strip tracker, a lead-tungstate crystal electromagnetic
calorimeter, and a brass/scintillator hadron calorimeter. Muons are
measured in gas-ionization detectors embedded in the
steel return yoke. Extensive forward calorimetry complements the
coverage provided by the barrel and endcap detectors.
Of particular importance to this search is the inner tracker~\cite{CMSTracker},
which consists of 1440 silicon pixel and 15\,148
silicon strip detector modules.
The inner tracker records 16 measurements on average per track.

The analysis is performed on the $\Pp\Pp$ collision data sample recorded by
the CMS detector
at $\sqrt{s} = 7\TeV$ in 2011, corresponding to an integrated
luminosity of 5.0\fbinv.

The data are selected with a single-muon trigger
that requires a track reconstructed in both the inner tracker and muon detectors
with transverse momentum $\pt > 40\GeV$
and pseudorapidity $|\eta|<2.1$, where
$\eta = - \ln [\tan(\theta/2)]$ and $\theta$ is the polar angle.
The radius of curvature of a fractionally charged particle in a magnetic
field is larger than
that of a particle of unit charge with the same momentum, so the
reconstructed $\pt$ is larger than the true
$\pt$ by the inverse of the particle's charge.
The trigger requirement that unit charge particles have $\pt >40\GeV$
corresponds to a requirement of
$\pt > 27$\GeV for
$L_{2/3}$ and $\pt > 13$\GeV for $L_{1/3}$.

The trigger efficiency for $L_{2/3}$ is in the range
67--74\% per event, which is very similar to the efficiency for unit charge
particles simulated with the same mass.
By contrast, the $L_{1/3}$ trigger efficiency is between 8\% and 18\%.
The lower trigger efficiency for $L_{1/3}$
results from the fact that many of the energy deposits in the tracker and
muon detectors are below the threshold required to record a
measurement.
The trigger efficiency for $L_{1/3}$ depends on the
particle's velocity $\beta$, since
\dedx $\propto 1/\beta^2$~\cite{PDG2012DeDx}.  The reconstruction efficiency for slower-moving
$L_{1/3}$ particles is larger because the energy deposits are more likely
to be above threshold.
As a result, the reconstructed velocity
distribution is very different from the generated distribution.  For
$L_{2/3}$ the larger overall efficiency means that
the velocity distribution is less affected by the reconstruction, but
slower moving $L_{2/3}$ particles fail the signal region requirement since
their recorded \dedx measurements are too large.

\section{Selection}
In a pre-selection step, candidates for fractionally charged particles are
defined as tracks reconstructed in the inner tracker and matched to a
track in the muon detectors.
The pre-selection criteria, which are described below, are chosen
to obtain well-reconstructed tracks and
to suppress background from cosmic ray muons.
We consider candidate muon tracks with large reconstructed transverse momenta,
$\pt>45\GeV$, as measured in the inner tracker,
in the range $|\eta| < 1.5$.
A loose requirement on the track fit quality,
$\chi^2/\mathrm{dof}<10$, rejects very poorly reconstructed tracks.
We also require at least six \dedx measurements from the
tracker, where a \dedx measurement is the signal amplitude recorded in
an inner detector module divided by the
particle's path length through the module.
To ensure that the track is isolated, the sum of the $\pt$ of all
other tracks within
a cone of $\Delta R \equiv \sqrt{(\Delta\phi)^2 + (\Delta\eta)^2}
<0.3$, where $\phi$ is the azimuthal angle,
around the candidate must be less than 0.1 times the
$\pt$ of the candidate.  The sum of the
electromagnetic and hadronic calorimeter energy recorded within this
cone, including that deposited by the candidate, must be less than 5\GeV.

Muons from cosmic rays are found in only a
small fraction of the triggered events, but because
they typically arrive out of coincidence with the bunch crossing,
the charge is sampled away from the signal's maximum,
and the resulting low \dedx measurement can be indistinguishable from
that of a fractionally charged particle.
Several requirements help to suppress the cosmic ray background.
The primary vertex with the smallest distance to the point of closest
approach of the candidate track is required to be a
well reconstructed vertex \cite{cms:vertexQuality}.
The track must also have at least two \dedx measurements in the pixel tracker.
The candidate track is required to
satisfy $|d_z|<0.5$\cm and $|d_{xy}|<0.1$\cm, where $d_z$ and $d_{xy}$
are the longitudinal and transverse impact parameters with respect to
the primary vertex.
Based on the time of flight measurement~\cite{HSCP2011} by the muon
detectors, we determine the time the candidate particle was
at the interaction
point (IP) under the assumptions that the particle had velocity
$\beta=1$ and
was moving outward from the IP.  This time must not be earlier than
the nominal bunch crossing that triggered the event, a requirement that
rejects over half of the background from $\Pp\Pp$ collisions and cosmic
rays.  We require $\alpha_\text{max} < 2.8\rad$, where
$\alpha_\text{max}$ is the maximum 3-dimensional opening angle between
the candidate track and any high-momentum ($\pt > 35\GeV$) track
reconstructed in either the inner tracker or in the muon system
alone.

The combined efficiency of the trigger and the pre-selection for
signal events generated with mass 100\GeV is
45\% per event for $L_{2/3}$ and 4.4\% for $L_{1/3}$.  The largest efficiency loss
is at the trigger stage for $L_{1/3}$, where a signal track may fail to be
reconstructed.  The pre-selection requirements reduce the signal
efficiency by roughly a factor of two.

After the pre-selection, events with two $L_q$ candidates are
rejected from the search sample if the invariant mass of the two
candidates $m_{LL}$ is in the range $80 < m_{LL} < 100\GeV$.  This
ensures that the search sample is independent of a
control sample, defined below, that is used to estimate the collisions
background.
Although in the considered signal model, $L_q \overline{L}_q$ are
produced in pairs, events containing a single candidate track are retained
after the pre-selection.
Events containing more than two candidates, which make up
less than 0.1\% of the search sample, are excluded.

We isolate the signal using a technique that imposes few assumptions
on any particular model but nonetheless has the power to suppress
the large standard model backgrounds from $\Pp\Pp$ collisions.
A fractionally charged particle is most clearly distinguished from a
standard model particle by its lower rate of energy loss in the
detector
since \dedx $\propto q^2$~\cite{PDG2012DeDx}.
Figure~\ref{fig:DeDxHitsDataVsCtrl}
shows the
distribution of \dedx measurements
associated with tracks passing the pre-selection, for the search
sample, a control sample, and the 100\GeV
$L_{2/3}$ and $L_{1/3}$ simulated signal samples.
The measured values from data lie
predominantly in the region above $2\MeVcm$.  We therefore define
low-ionizing
measurements to be those with $\dedx<2\MeVcm$.  By requiring a
number of such low \dedx measurements, standard model backgrounds
can be suppressed while most of the signal events that pass the
pre-selection are retained.
Tracks that intersect a sensor close to its edge or near the boundary
between two sensor modules can result in low \dedx measurements
because of the partial collection of the deposited charge, so these
track measurements are
not considered in the analysis.  The distance to the sensor edge
for which \dedx measurements are excluded is between 0.5\% and 5\% of the distance
to the center of the module, depending on the
tracker subsystem.

A signal region is determined by
maximizing the
expected mass limit on the
production cross section of fractionally charged particles
while varying the minimum number of
low \dedx measurements.  The signal region optimization is
performed for simulated samples with a $L_q$ mass of $100\GeV$ and
$400\GeV$, and for both samples the optimum signal region is
defined with the requirement that events contain a track with
at least six low \dedx measurements.  For the 100\GeV signal events
passing the
pre-selection, 75\% of $L_{2/3}$ events are in the signal region,
and 93\% of $L_{1/3}$ events are in the signal region.

\section{Background estimate}
We use control samples of data to estimate the background contribution
from cosmic ray muons and from particles
produced in $\Pp\Pp$ collisions.
The data-driven method provides a background estimate without the use of
simulation.

To estimate the cosmic ray background, we use a sample of muons
obtained with the nominal pre-selection except
for two inverted requirements, $0.1 < |d_{xy}| < 1.1$\cm and
$0.5 < |d_{z}| < 50$\cm.
The yield in the signal region of this sample is scaled by a weight
factor to obtain the cosmic ray background for the nominal pre-selection.
This weight factor is the product of two weights,
each determined from a cosmic ray enriched sample as the ratio of the
yield in the nominal selection region to the yield after inverting a
single requirement.
This scaled yield gives a cosmic ray background estimate of $0.007$ events.

\begin{figure}[!Hhtb]
   \centering
    \includegraphics[width=\cmsFigWidth]{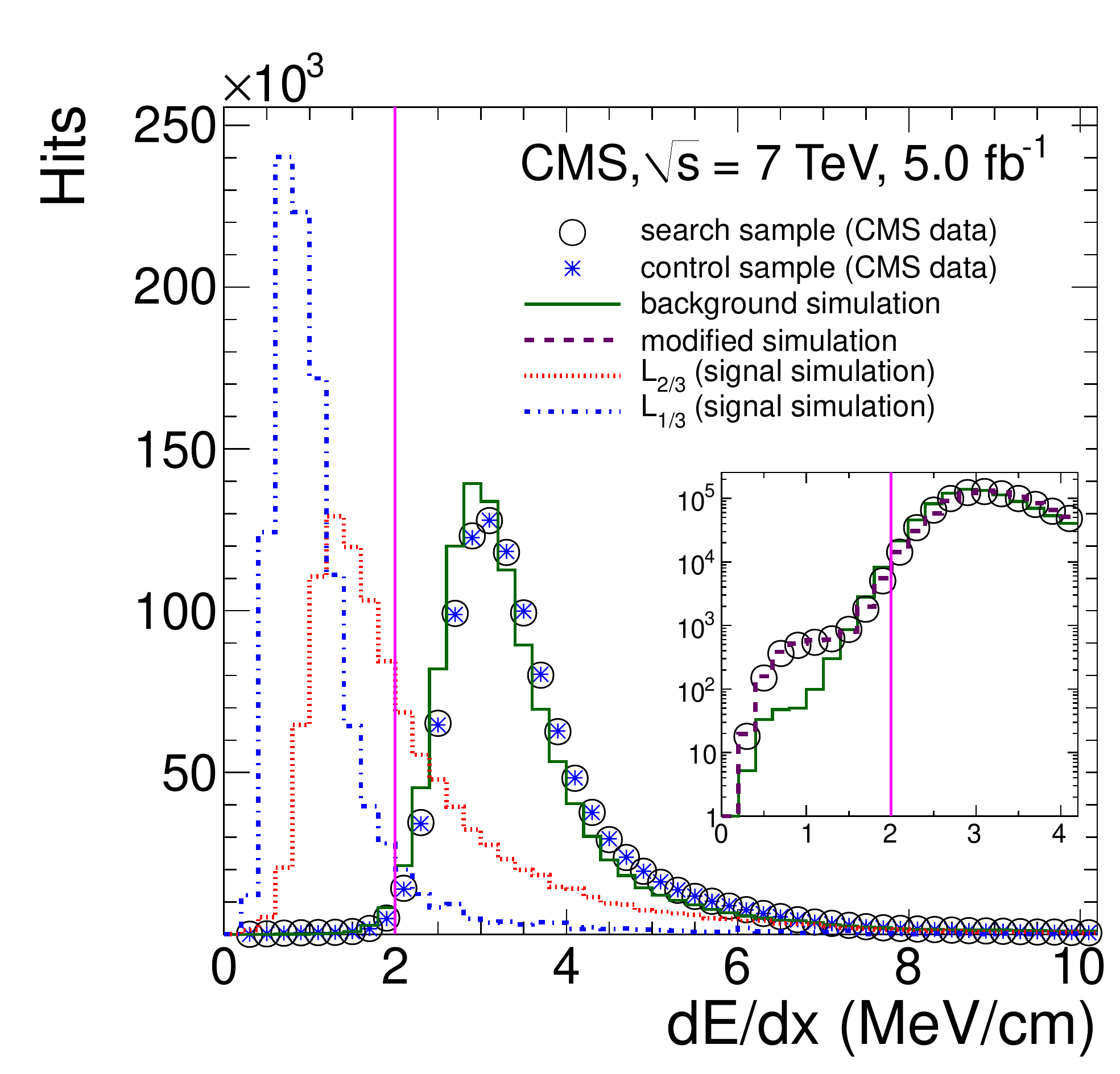}
  \caption[\dedx, search data versus control sample]{
Distribution of \dedx measurements associated with tracks passing the
pre-selection in the search sample and the \cPZ-peak control sample.
Simulated $L_q \overline{L}_q$ signal samples for a mass
of $100\GeV$ are shown for $q=$ 2/3, 1/3.
The distributions
are normalized to the area of the
search sample.
The magenta vertical line at $\dedx=2\MeVcm$ indicates the upper limit of
the range of measurements considered to be low-ionizing.
The inset is an enlargement of the region of low \dedx, plotted on a
semi-logarithmic scale.
\label{fig:DeDxHitsDataVsCtrl}
}
\end{figure}

A \cPZ-peak control sample is used to estimate the $\Pp\Pp$ collision
background.  This sample is selected by relaxing the transverse momentum
requirement to $\pt>35 \GeV$, requiring $80 < m_{LL} < 100\GeV$, and
applying all other pre-selection requirements.
Figure~\ref{fig:DeDxHitsDataVsCtrl} shows the distributions of \dedx
measurements associated with
selected tracks for both the search sample and the \cPZ-peak control
sample. The two distributions agree within the statistical
uncertainties over the full \dedx range.
The ratio of the number
of tracks passing the pre-selection in the search sample to the number
of tracks passing the pre-selection in the \cPZ-peak control sample is 10.5.
The control sample is scaled by this ratio to model the distribution
in the search sample.

The simulation of the control sample
is also shown in Fig.~\ref{fig:DeDxHitsDataVsCtrl}. This simulation is used
only to assess the uncertainty in the signal
efficiency, since the background estimate is entirely data-driven.
The inset in Fig.~\ref{fig:DeDxHitsDataVsCtrl} is an enlargement of
the region of low \dedx, plotted on a semi-logarithmic scale.
This inset also shows the results of a modified simulation, which
includes the effect of a possible source of anomalously low \dedx hits
not reproduced in the nominal simulation.  The background simulations
are discussed in the next section.

\begin{figure}[!Hhtb]
    \centering
   \includegraphics[width=\cmsFigWidth]{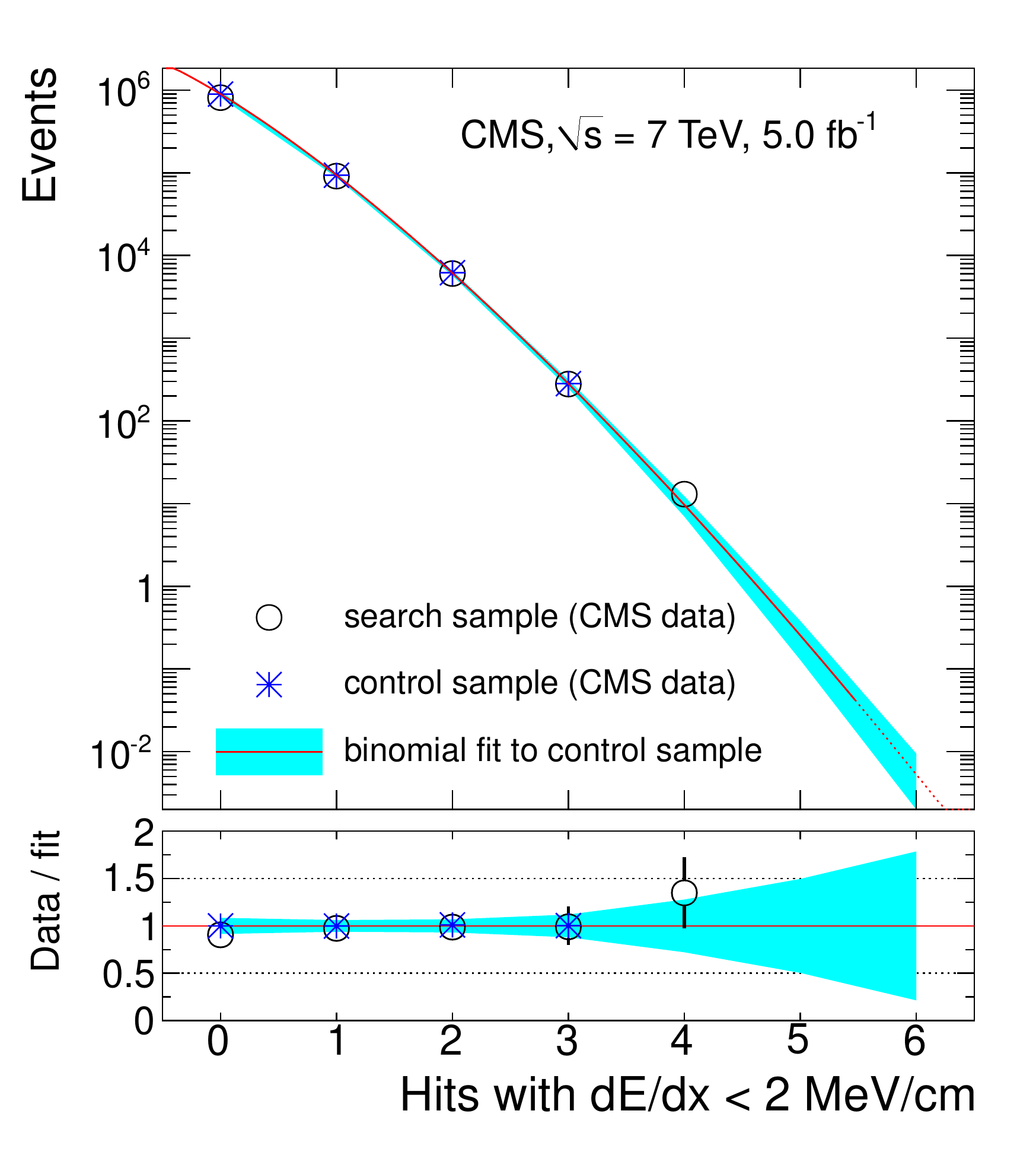}
  \caption[Number of low-\dedx hits]{Number of events with at least one track with the given number of
low \dedx measurements, for search data and the scaled $\cPZ$-peak control sample
background estimate.
The binomial function fit to the control sample is shown, with the
band representing its uncertainty.  The ratio of the data to
the binomial fit is also presented.  No tracks in the search sample
have five or more low \dedx measurements.
  \label{fig:NLoHitsDataVsEst}
}
\end{figure}

To estimate the background in the signal region, we extrapolate from
the background-dominated region of
events containing a pre-selected track with zero to five low \dedx
measurements.
For a muon from a $\cPZ$ decay, the \dedx measurements associated
with the track are expected to be uncorrelated, and
the number of measurements
below a given \dedx value can be described by a generalized binomial
function,
\begin{eqnarray}
N_\text{evts} = N_0 \binom{\mu}{n} p^n (1-p)^{\mu - n} , \nonumber \\
\binom{\mu}{n} =
\frac{\Gamma(\mu + 1)}{
\Gamma(n+1)
\Gamma(\mu - n + 1)} , \nonumber
\end{eqnarray}
where $N_\text{evts}$ is the number of
events containing at least one track with $n$ low \dedx measurements,
$\mu$ is the average number of measurements per track,
$p$ is the probability for a single measurement to be low \dedx,
$N_0$ is a normalization factor, and $\Gamma(n)$ is the gamma function.
The fit of the binomial function to the
background-dominated region of the $\cPZ$-peak control sample is shown in
Fig.~\ref{fig:NLoHitsDataVsEst}.
The fitted parameters are
$\mu = 17.5 \pm 1.7$,
$p = 0.0125 \pm 0.0013$, and
$N_0 = (5.03 \pm 0.03) \times 10^6$; the values of $\mu$ and $p$ are
close to those expected based on the number of measurements per track
and the fraction of low \dedx measurements.
This function describes the distribution in the
control sample well, with $\chi^2/\mathrm{dof} = 0.07/1$,
corresponding to a $\chi^2$ probability of 79\%.
This is strong support for the hypothesis that the data are
distributed binomially
and therefore that the \dedx measurements are uncorrelated.
Extrapolation of the fitted binomial function into
the signal region yields a $\Pp\Pp$ background estimate of
$0.005$ events.

\section{Systematic uncertainties}
The systematic uncertainties that
significantly impact the results are the
uncertainties in the integrated luminosity,
the background estimate, and the signal efficiency.
The uncertainty in the integrated luminosity is 2.2\%~\cite{lumi2011}.

The cosmic ray background estimate has a statistical uncertainty of
71\% that arises from the relatively small size of the sample with inverted
$d_{xy}$ and $d_{z}$ requirements used for its determination.
The statistical uncertainties in the weighting factors are 1\% and 24\% for the $d_{xy}$
and $d_{z}$ requirements, respectively.  The systematic uncertainty associated with the assumption that the
$d_{xy}$ and $d_{z}$ variables are uncorrelated is assessed by
examining a sample defined by replacing the inverted $d_z$ selection
with an inverted $\alpha_\text{max}$ requirement.  This sample, obtained by requiring
$0.1 < |d_{xy}| < 1.1\unit{cm}$, $\alpha_\text{max} > 2.8\unit{rad}$, and all other
pre-selection criteria, provides a second estimate of the comsic ray
background, which differs from the nominal estimate by 42\%.
The statistical and systematic uncertainties are summed in quadrature;
the total cosmic ray background estimate is $0.007 \pm 0.006$ events.

We assess three potential sources of uncertainty in the $\Pp\Pp$ background estimate in the
signal region. The first source is from the choice of the function used to
fit the control sample.  While this is often a large source of
uncertainty in many a posteriori fits to data, our hypothesis that a
binomial function describes the distribution of the number of low
\dedx measurements is motivated a priori, from first principles.   We do
not expect a large uncertainty from this source.  For completeness,
however, other functions are also compared to the data.
Fits of several modified exponential, power-law, and polynomial
functions fail to converge or have very low $\chi^2$ probabilities.
One function that does fit the distribution reasonably well is
$N_\text{evts} = p_0 n^{p_1 + p_2 n}$, where $p_i$ are free
parameters.  The difference between the background estimate from
this function and nominal background estimate is 0.001 events.

The second potential source of uncertainty in the $\Pp\Pp$ background
estimate arises from the statistical uncertainties in the fitted
parameters of the binomial function.
The propagation of these
uncertainties results in an uncertainty in the
background estimate of ${\pm}0.0004$ events.

A third source of uncertainty arises from the small disagreement
between the distribution of low \dedx
measurements from the control sample and that from the search sample.
In the background-dominated region,
the largest statistically-significant discrepancy between the two
samples is 9\%, for zero low \dedx measurements.  To assess the resulting systematic uncertainty, the control sample fit is
repeated for a large number of trials, in each case setting the value
of the distribution in each bin randomly, according to a Gaussian distribution with a mean of the
original value and width of 9\% of the original value.  The RMS of the background estimates from all of these trials is
0.004 events, which is taken as the uncertainty due to the accuracy with which the control sample models
the search sample.

The three sources of uncertainty in the $\Pp\Pp$ background estimate are summed in
quadrature giving a total estimate of $0.005 \pm 0.004$ events.
The high precision of this estimate is due to the large
statistics of the control sample, which leads to small statistical
uncertainties in the fitted parameters.  Likewise, the high degree of
accuracy of the background estimate is reflected in the relatively small
systematic uncertainty assigned.  This is a direct consequence of the
somewhat unusual aspect of this analysis that the functional form with
which the data would be distributed under the background only
hypothesis was derived from first principles.

The sum of the $\Pp\Pp$ and cosmic ray backgrounds gives a total
background estimate of $0.012 \pm0.007$ events.

This search uses the data itself to estimate all backgrounds, so the
uncertainties in the simulation only impact the determination of the
efficiency of the benchmark signal model.
The systematic uncertainties in the signal efficiencies are summarized
in Table \ref{tab:systSumm}.

The simulation of the trigger efficiency
for a fractionally charged particle is sensitive to the accurate
modeling of the muon detectors' electronics and gas gain as well as
the threshold for recording a hit, because it has less ionization
energy loss and thus the peak of its Landau
distribution is closer to the threshold than that of a muon.
We assess the impact on the signal region efficiency of a variation in
the simulated gain of the muon system by a conservative estimate of
its uncertainty.
The impact of such a variation on $L_{2/3}$ is
small, since the charge distributions are typically far above the
threshold.  However, for $L_{1/3}$, the probability to record a
hit degrades significantly as the gain decreases.  The impact of this
variation for $L_{2/3}$ is 1\%, and for $L_{1/3}$ is 8.5\%.
The systematic uncertainty in the offline global muon
identification requirement is negligible by comparison.

The modeling of the tracker \dedx measurements impacts the simulated signal
efficiency by affecting the efficiency of track reconstruction and
signal region selection.
Larger tracker energy deposits are more
likely to be above the threshold required to record a measurement, and
the track reconstruction efficiency increases with more measurements.
Larger \dedx measurements also reduce the fraction of reconstructed
tracks that are in the signal region, since fewer measurements are
below the $2\MeVcm$ limit.
To evaluate the accuracy of the simulation of the \dedx measurements,
we compare the \dedx distributions of the \cPZ-peak control sample in
simulation and in data, as shown in Fig.~\ref{fig:DeDxHitsDataVsCtrl}.
The agreement in the low-\dedx region is evaluated as
the shift of all \dedx measurements required to obtain the same fraction below
$2\MeVcm$ in both samples.
For the nominal
selection, the simulated \dedx distribution must be shifted by 2.6\% to
obtain the same fraction below $2\MeVcm$ as in the data.  For a larger
sample obtained with looser
selection criteria, the required shift is 5\%.
We use the larger of these values, 5\%, as an estimate of
the agreement between simulation and data.  To assess the resulting
uncertainty in the signal efficiency,
we vary the amplitude of the \dedx measurements
by ${\pm}5\%$ before re-simulating the trigger emulation, track
reconstruction, and full selection.
A variation of $+5\%$ in the \dedx measurements changes the
signal efficiency by $+16\%$ for $L_{1/3}$ particles and
$-7.5\%$ for $L_{2/3}$.
The efficiency changes in opposite directions because for $L_{1/3}$
the increased reconstruction efficiency is the
dominant effect, while for $L_{2/3}$ the smaller signal region
efficiency has a greater impact.
These variations in the
signal efficiency are taken as the systematic uncertainties associated with the
modeling of the tracker \dedx measurements.

Potential causes of incorrect modeling of \dedx in our simulation that could
produce a disagreement at low \dedx have been examined. The most
likely possibility is a residual source of low \dedx hits
that are not removed by the sensor-edge fiducial cuts.
Such a source could be accommodated by the control sample data if, at
most, 0.06\% of all measurements are affected. A simulation
assuming a mismeasurement rate at this level reproduces the observed
data distribution, as shown in the inset of
Fig.~\ref{fig:DeDxHitsDataVsCtrl}. Such an effect would impact less
than 1\% of all tracks and
change the signal efficiency by an even smaller amount. Furthermore,
the likelihood of a track to have six such anomalous
measurements is extremely small, so the effect on the background
estimate would be negligible.

The uncertainty associated with the track momentum scale is less than 1\%.
The impact of the uncertainty in the muon timing measurements is 2\%.
This is assessed by varying the timing measurements according to the
measured discrepancy between the data and simulation, as described
in~\cite{HSCP2011}.

\begin{table}[htb]
\topcaption
{Systematic uncertainties (in \%) in the signal efficiency.
}
\scotchrule[lcc]
Source                     & $L_{1/3}$  & $L_{2/3}$  \\
\hline
Muon trigger	 	   &    8.5 &   1 	\\
Tracker \dedx measurements &   16   &   7.5 	\\
Track momentum scale	   &  ${<}1$  & ${<}1$ 	\\
Muon timing                &    2   &   2       \\
\hline
\textbf{Total} 	           &   18   &   8 	\\
\donescotchrule
\label{tab:systSumm}
\end{table}

\section{Results}
\begin{figure}[hbtp]
  \centering
  \includegraphics[width=\cmsFigWidth]{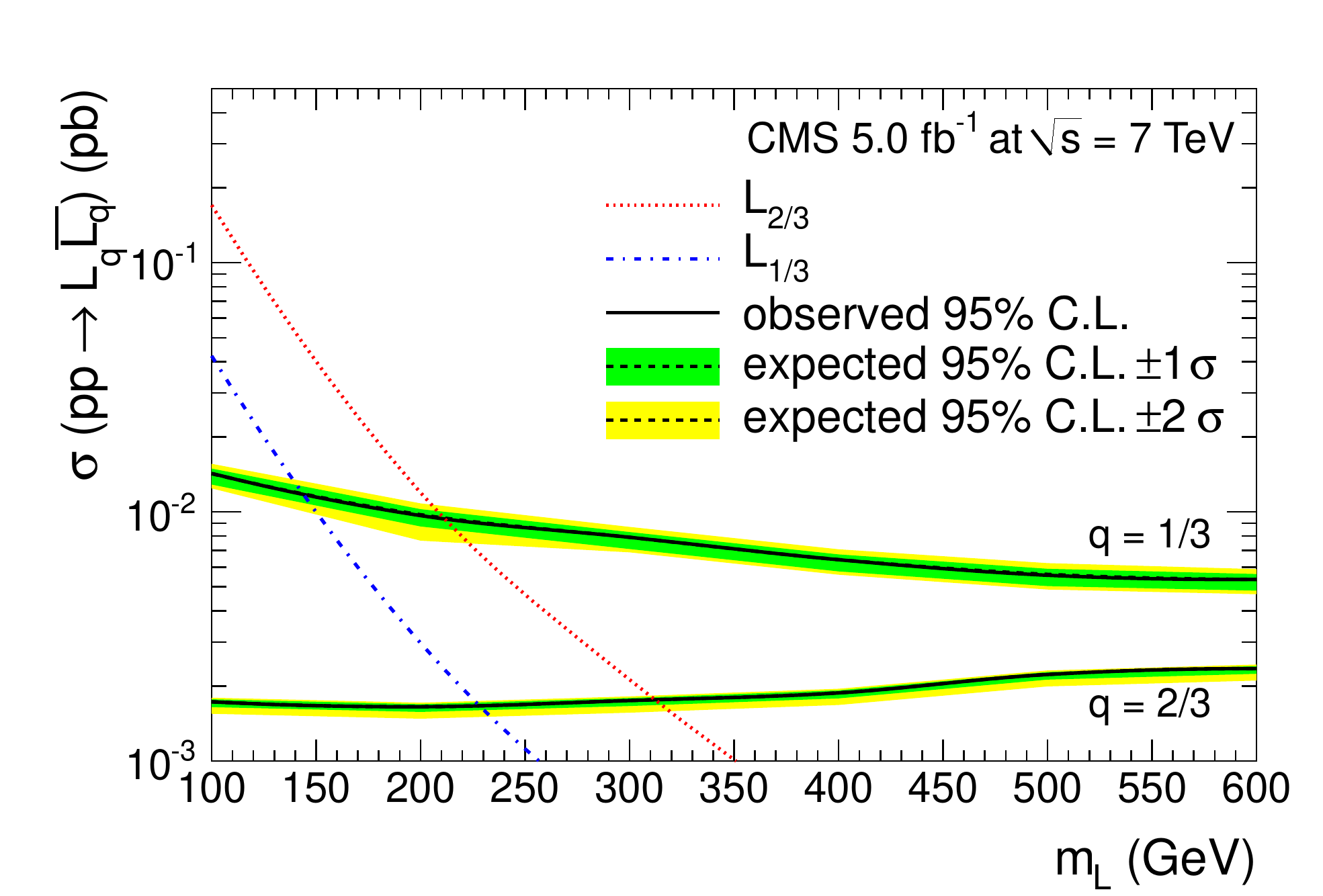}
  \caption{Expected and observed limits on the cross section of
    $L_q\overline{L}_q$ for $q=2/3$ and 1/3.
The theoretical prediction of modified Drell--Yan production
with $t_{3L}$ = 0 is shown.
 The lines for the expected and observed limits are overlapping.
  \label{fig:limits}
}
\end{figure}

The numbers of expected background and observed events are summarized in Table
\ref{tab:bkgSummary}.  We observe zero events in the data search sample in the signal
region, which is consistent with the background estimate of $0.012 \pm 0.007$
events.  The  $L_q \overline{L}_q$ signal efficiency and average
$\beta$ for different signal masses and charges are listed in
Table \ref{tab:limits}. Ninety-five percent confidence level (CL) upper
limits on the $L_q\overline{L}_q$ production cross section are
calculated using the CL$_\mathrm{s}$ criterion \cite{ReadCLs, JunkCLs}.  Expected
and observed 95\% CL limits are shown in Fig.~\ref{fig:limits}.
These limits vary from 1.7 to 2.3\unit{fb}, for $q=2/3$, and from 14 to 5.4\unit{fb},
for $q=1/3$, for masses between 100 and 600\GeV.
We exclude the production of $L_{2/3}$
with a mass below 310\GeV and the production of $L_{1/3}$
with a mass below 140\GeV at 95\% CL.
 \begin{table}
    \centering
    \topcaption{Numbers of events in the signal region, containing a track
      with at least six \dedx measurements with $\dedx< 2\MeVcm$, estimated from the
      background and observed in the search
      data.    \label{tab:bkgSummary}
    }
    \scotchrule[lc]
      Cosmic rays       & $0.007 \pm 0.006$   \\
      $\Pp\Pp$ collisions     & $0.005 \pm 0.004$  \\
      Total background  & $0.012 \pm 0.007$ \\
      \hline
      Observed events   & 0 \\
    \donescotchrule
  \end{table}
   \begin{table}[htb]
 \topcaption
 {The signal efficiency and
   average velocity $\langle\beta\rangle$ of events in the signal region for different mass points
   and charge hypotheses. \label{tab:limits}
}
   \scotchrule[ccc|cc]
 Mass         & \multicolumn{2}{c|}{$L_{2/3}$}       & \multicolumn{2}{c}{$L_{1/3}$}  \\
 (\gev)       & Signal eff. & $\langle\beta\rangle$ & Signal eff. & $\langle\beta\rangle$  \\
 \hline

100  & 0.341 $\pm$ 0.026  & 0.84  & 0.041 $\pm$ 0.007  & 0.52  \\
200  & 0.357 $\pm$ 0.027  & 0.83  & 0.060 $\pm$ 0.011  & 0.51  \\
300  & 0.337 $\pm$ 0.026  & 0.82  & 0.074 $\pm$ 0.013  & 0.51  \\
400  & 0.314 $\pm$ 0.024  & 0.80  & 0.091 $\pm$ 0.016  & 0.51  \\
500  & 0.265 $\pm$ 0.020  & 0.79  & 0.104 $\pm$ 0.019  & 0.51  \\
600  & 0.251 $\pm$ 0.019  & 0.78  & 0.109 $\pm$ 0.019  & 0.51  \\
\donescotchrule
\end{table}

\section{Conclusion}

A search has been performed for heavy, long-lived,
lepton-like fractionally charged particles, using the signature of at
least six low \dedx measurements.
The search is based on a $\Pp\Pp$ collision sample recorded by
the CMS detector at $\sqrt{s} = 7\TeV$, corresponding to an integrated
luminosity of 5.0\fbinv.
Zero events are observed in the signal region, consistent with
the background estimate.
Upper limits on the production cross section of
pair produced, spin-1/2 particles that are neutral under $SU(3)_C$ and $SU(2)_L$
exclude at 95\% confidence $m_L$ below 310\GeV for particles with $q=2/3$ and
$m_L$ below 140\GeV for
particles with $q=1/3$.

\section{Acknowledgements}
We thank Nathaniel Craig for performing checks of the signal
model considered in this analysis.
We congratulate our colleagues in the CERN accelerator departments for
the excellent performance of the LHC machine. We thank the technical
and administrative staff at CERN and other CMS institutes, and
acknowledge support from: BMWF and FWF (Austria); FNRS and FWO
(Belgium); CNPq, CAPES, FAPERJ, and FAPESP (Brazil); MES (Bulgaria);
CERN; CAS, MoST, and NSFC (China); COLCIENCIAS (Colombia); MSES
(Croatia); RPF (Cyprus); MoER, SF0690030s09 and ERDF (Estonia);
Academy of Finland, MEC, and HIP (Finland); CEA and CNRS/IN2P3
(France); BMBF, DFG, and HGF (Germany); GSRT (Greece); OTKA and NKTH
(Hungary); DAE and DST (India); IPM (Iran); SFI (Ireland); INFN
(Italy); NRF and WCU (Korea); LAS (Lithuania); CINVESTAV, CONACYT,
SEP, and UASLP-FAI (Mexico); MSI (New Zealand); PAEC (Pakistan); MSHE
and NSC (Poland); FCT (Portugal); JINR (Armenia, Belarus, Georgia,
Ukraine, Uzbekistan); MON, RosAtom, RAS and RFBR (Russia); MSTD
(Serbia); SEIDI and CPAN (Spain); Swiss Funding Agencies
(Switzerland); NSC (Taipei); TUBITAK and TAEK (Turkey); STFC (United
Kingdom); DOE and NSF (USA).

\bibliography{auto_generated}   

\cleardoublepage \appendix\section{The CMS Collaboration \label{app:collab}}\begin{sloppypar}\hyphenpenalty=5000\widowpenalty=500\clubpenalty=5000\textbf{Yerevan Physics Institute,  Yerevan,  Armenia}\\*[0pt]
S.~Chatrchyan, V.~Khachatryan, A.M.~Sirunyan, A.~Tumasyan
\vskip\cmsinstskip
\textbf{Institut f\"{u}r Hochenergiephysik der OeAW,  Wien,  Austria}\\*[0pt]
W.~Adam, E.~Aguilo, T.~Bergauer, M.~Dragicevic, J.~Er\"{o}, C.~Fabjan\cmsAuthorMark{1}, M.~Friedl, R.~Fr\"{u}hwirth\cmsAuthorMark{1}, V.M.~Ghete, J.~Hammer, N.~H\"{o}rmann, J.~Hrubec, M.~Jeitler\cmsAuthorMark{1}, W.~Kiesenhofer, V.~Kn\"{u}nz, M.~Krammer\cmsAuthorMark{1}, I.~Kr\"{a}tschmer, D.~Liko, I.~Mikulec, M.~Pernicka$^{\textrm{\dag}}$, B.~Rahbaran, C.~Rohringer, H.~Rohringer, R.~Sch\"{o}fbeck, J.~Strauss, A.~Taurok, W.~Waltenberger, G.~Walzel, E.~Widl, C.-E.~Wulz\cmsAuthorMark{1}
% ...
\vskip\cmsinstskip
\dag:~Deceased\\
1:~~Also at Vienna University of Technology, Vienna, Austria\\
% ...

\end{sloppypar}
\end{document}